%% file: main.tex
\documentclass[envcountsect]{llncs}
\input{setup/packages}

\input{setup/macros}

\input{setup/table}

\begin{document}
\frontmatter
\mainmatter

\title{Not so immutable: Upgradeability of Smart Contracts on Ethereum}

\author{Mehdi Salehi \and Jeremy Clark \and Mohammad Mannan}
\institute{Concordia University}

\maketitle


\input{sections/abstract}


\input{sections/body}



\subsubsection*{Acknowledgements.} Our measurements were possible thanks to \url{https://archivenode.io/}. We thank Santiago Palladino (OpenZeppelin) and the reviewers for comments and discussions that helped to improve our paper. J. Clark acknowledges support for this research project from the National Sciences and Engineering Research Council (NSERC), Raymond Chabot Grant Thornton, and Catallaxy Industrial Research Chair in Blockchain Technologies and the AMF (Autorité des Marchés Financiers). J. Clark and M. Mannan acknowledge NSERC through Discovery Grants.


\bibliography{bib/pulp.bib,bib/new.bib}


\clearpage
\appendix

\input{sections/appendix}


\end{document}

%% file: setup/packages.tex
\usepackage[pdftex]{graphicx}
\graphicspath{{figures/}}
\DeclareGraphicsExtensions{.jpg,.png}
\usepackage[table,xcdraw]{xcolor}

\usepackage[caption=false,font=footnotesize]{subfig}

\usepackage{amsmath}
\usepackage{bbm}
\usepackage{stmaryrd}
\usepackage{wasysym}
\usepackage{amssymb}
\usepackage{color}
\usepackage{multirow}
\usepackage{rotating}
\usepackage{makecell}
\usepackage{hhline}
\usepackage{array}
\usepackage{color}
\usepackage[hyphens]{url}
\usepackage[pdftitle=Title,pdfauthor=Anonymous]{hyperref}
\usepackage{comment}

%% file: setup/macros.tex
\usepackage{xspace}

\newcommand{\etc}{\textit{etc.}\xspace}
\newcommand{\ie}{\textit{i.e.,}\xspace}
\newcommand{\eg}{\textit{e.g.,}\xspace}



%% file: setup/table.tex
\usepackage{adjustbox}
\newcommand{\headrow}[1]{\multicolumn{1}{c}{\adjustbox{angle=30,lap=\width-0.5em}{#1}}}

\newcommand{\full}{$\bullet$}
\newcommand{\prt}{$\circ$}

%% file: sections/abstract.tex

\begin{abstract}
A smart contract that is deployed to a blockchain system like Ethereum is, under reasonable circumstances, expected to be immutable and tamper-proof. This is both a feature (promoting integrity and transparency) and a bug (preventing security patches and feature updates). Modern smart contracts use software tricks to enable upgradeability, raising the research questions of \textit{how} upgradeability is achieved and \textit{who} is authorized to make changes. In this paper, we summarize and evaluate six upgradeability patterns. We develop a measurement framework for finding how many upgradeable contracts are on Ethereum that use certain prominent upgrade patters. We find 1.4 million proxy contracts which \textit{8,225} of them are unique upgradeable proxy contracts. We also measure how they implement access control over their upgradeability: about 50\% are controlled by a single Externally Owned Address (EOA), and about 14\% are controlled by multi-signature wallets in which a limited number of persons can change the whole logic of the contract.

\end{abstract}

%% file: sections/body.tex


\section{Introductory Remarks}

The key promise of a smart contract running on Ethereum is that its code will execute exactly as it is written, and the code that is written can never be changed. While Ethereum cannot maintain this promise unconditionally, its assumptions (\eg cryptographic primitives are secure and well-intentioned participants outweigh malicious ones) provide a realistic level of assurance. 

The immutability of a smart contract's code is related to trust. If Alice can validate the code of a contract, she can trust her money to it and not be surprised by its behavior. Unfortunately, disguising malicious behavior in innocuous-looking code is possible (`rug pulls'), and many blockchain users have been victims. On the other hand, if the smart contract is long-standing with lots of attention, and security assessments from third-party professional auditors, the immutability of the code can add confidence. 

The flip-side of immutability is that it prevents software updates. Consider the case where a security vulnerability in the code of a smart contract is discovered. Less urgently, some software projects may want to roll out new features, which is also blocked by immutability. There is an intense debate about whether this is a positive or negative, with many claiming that `upgradeability is a bug.'~\footnote{\href{https://medium.com/consensys-diligence/upgradeability-is-a-bug-dba0203152ce}{``Upgradeability Is a Bug'', Steve Marx, Medium, Feb 2019.}} We do not take a position on this debate. We note that upgradeability is happening and we seek to study what is already being done and what is possible. 

Is there a way to deploy upgradeable smart contracts if all smart contracts are (practically speaking) immutable? Consider two simple ideas. The first is to deploy the upgraded smart contract at a new address. One main drawback to this is that all software and websites need to update their addresses. A second simple idea is to use a proxy contract (call it P) that stores the address of the `real' contract (call it A). Users consider the system to be deployed at P (and might not even be aware it is proxy). When a function is called on P, it is forwarded to A. When an upgrade is deployed to a new address (call it B), the address in P is changed from A to B. This solution also has drawbacks. For example, if the proxy contract hardcodes the list of functions that might be called on A, new functions cannot be added to B. Another issue is that the data (contract state) is stored in A. For most applications, a snapshot of A's state will need to be copied to B without creating race conditions. Mitigating these issues leads to more elaborate solutions like splitting up a contract logic and state, utilizing Ethereum-specific tricks (fallback functions to capture unexpected function names), and trying to reduce the gas costs of indirection between contracts.

\paragraph{Contributions and Related Work.} 
The state of smart contract upgradeability methods in Ethereum is mainly discussed in non-academic, technical blog posts~\cite{openzeppelinPost,tobBlogPost}. In Section~\ref{sec:classification}, we systemize the different types using these resources, and provide a novel evaluation framework for comparing them.

Fr{\"o}wis and B{\"o}hme~\cite{frowisnot} conducted a measurement study on the use-cases of the \texttt{CREATE2} opcode in Ethereum blockchain, which one of them is the Metamorphosis upgradeability pattern discussed in Section~\ref{sec:metamorphic}. They also find, in a passing footnote, some delegate-call based contracts by assuming compliance with the standards: EIP-897, EIP-1167, EIP-1822, and EIP-1967. In our paper, we contribute a more general pattern-based measurement that is not specific to a standard or a commonly-used implementation. We also are the first, to our knowledge, to study who is authorized for upgrading upgradeable contracts, shedding light on the risks of different admin types.

Recent papers have provided security tools for developers that compose with upgradeablity patterns based on  \texttt{DELEGATECALL}~\cite{rodler2021evmpatch,perez2022dissimilar}. Numerous measurement studies have used Ethereum blockchain data but concern aspects other than upgradeability~\cite{perez2019broken,chen2017adaptive,reijsbergen2021transaction,victor2019measuring,pinna2019massive,he2020characterizing}. Chen et al.~\cite{chen2021smart} survey use-cases of the \textit{SELFDESTRUCT} opcode, but they do not cover how it is used in Metamorphosis~\ref{sec:metamorphic}.

\section{Classification of Upgrade Patterns} \label{sec:classification}

\paragraph{Updating vs. upgrading.} Software maintenance is part of software's lifecycle, and the process of changing the product after delivery. Often a distinction is drawn between software \textit{updates} and software \textit{upgrades}. An update modifies isolated portions of the software to fix bugs and vulnerabilities. An upgrade is generally a larger overhaul of the software with significant changes to features and capabilities. We only use the term upgrade and distinguish between retail (parameters and isolated code) and wholesale (entire application) changes to a smart contract. While upgrades to a smart contract's user interface (UI) can significantly change a user experience and expose new features, UIs are governed by traditional software maintenance. Our paper only considers the on-chain smart contract component, which is significantly more challenging to upgrade as it is on-chain and immutable under reasonable circumstances.

\begin{figure}[t]
  \centering
      \includegraphics[width=0.8\textwidth]{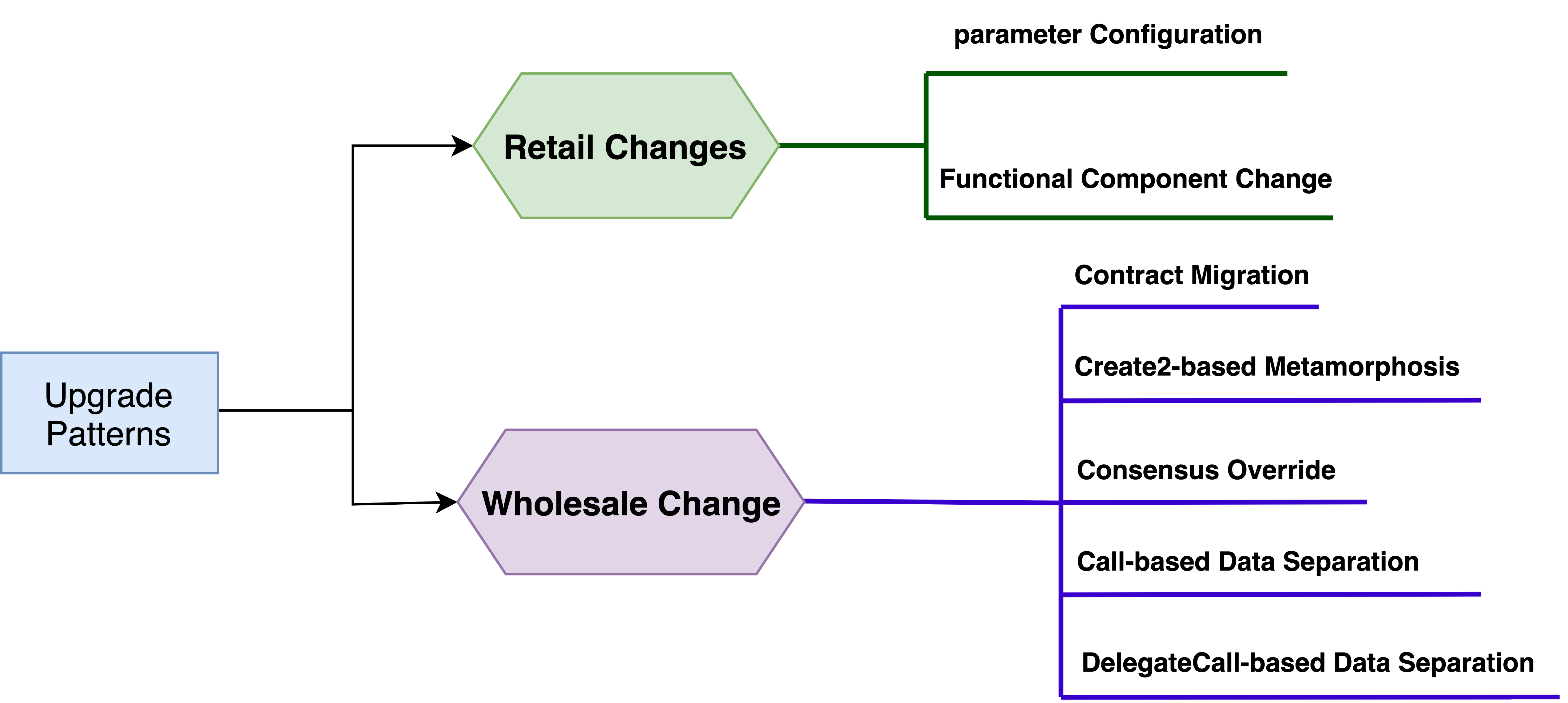}
  \caption{Classification of upgradeability patterns.\label{fig:class}}
 \end{figure}
 
A variety of upgradeability patterns have been proposed for smart contracts. Most leverage Ethereum-specific operations and memory layouts and are not applicable to other blockchain systems.




\subsection{Parameter Configuration}
\label{sec:parameter}

We first categorize upgradeability patterns into two main classes: \textit{retail changes} and \textit{wholesale changes}. A pattern for retail change does not enable the replacement of the entire contract. Rather, a component of the contract is pre-determined (before the contract is deployed on Ethereum) to allow future upgrades, and the code is adjusted to allow these changes. 

The simplest upgrade pattern is to allow a system parameter, that is stored in a state variable, to be changed. This requires a \textit{setter function} to overwrite (or otherwise adjust) the variable, and access control over who can invoke the function. For example, in decentralized finance (DeFi), many services have parameters that control fees, interest rates, liquidation levels, \etc Adjustments to these parameters can initiate large changes in how the service is used (its `tokenomics'). A DeFi provider can retain control over these parameters, democratize control to a set of token holders (\eg stability fees in the stablecoin project MakerDao), or lock the parameters from anyone's control. In Section~\ref{sec:governance}, we dive deeper into the question who can upgrade a contract. 


\subsection{Functional Component Change}
\label{sec:component}

While a parameter change allows an authorized user to overwrite memory, a functional component change addresses modifications to the code of a function (and thus, the logic of the contract). In the EVM, code cannot be modified once written and so new code must be deployed to a new contract, but can be arranged to be called from the original contract. 

One way to allow upgradable functions is deploying a helper contract that contains the code for the functions to be upgradeable. Users are given the address of the primary contract, and the address of this secondary contract is stored as a variable in the primary contract. Whenever this function is invoked at the primary contract, the primary contract is pre-programmed to forward the function call, using the opcode \texttt{Call}, to the address it has stored for the secondary contract. To modify the logic of the function, a new secondary contract is deployed at a new address, and an authorized set of individuals can then use a parameter change in the primary contract to update the address of the secondary contract.

The DeFi lending platform Compound~\footnote{\url{https://compound.finance}} uses this pattern for their interest rate models~\footnote{\url{https://github.com/compound-finance/compound-protocol/blob/v2.3/contracts/InterestRateModel.sol}} which are tailored specifically for each asset. The model for one asset can be changed without impacting the rest of the contract~\cite{openzeppelinPost}.

Upgradeable functional components need to be pre-determined before deploying the primary contract. Once the primary contract is deployed, it is not possible to add upgradeability to existing functions. It also cannot be directly used to add new functions to a contract. Finally, this pattern is most straightforward when the primary contract only uses the return value from the function to modify its own state. Thus, the function is either `pure' (relies only on the parameters to determine the output) or `view' (can read state from itself or other contracts, but cannot write state). If the function modifies the state of the primary contract, the primary contract must either expose its state variables to the secondary contract (by implementing setter functions), or it can run the function using \texttt{Delgatecall} if the secondary contract has no state of its own. 

This upgrade pattern suggests a way forward for wholesale changes to the entire contract: create a generic `proxy' contract that forwards all functions to a secondary contract. To work seamlessly, this requires some further engineering (Sections~\ref{sec:callbased} and \ref{sec:delegatecall}).





\subsection{Consensus Override}
\label{sec:hardfork}

The two previous patterns enable portions of a smart contract to be modified. The remaining patterns strive to allow an entire contract to be modified or, more simply, replaced. The first wholesale pattern is not a tenable solution to upgradeability as it as only been used rarely under extraordinary circumstances, but we include it for completeness. 

Immutability is enforced by the consensus of the blockchain network. If participating nodes (\eg miners) agreed to suspend immutability, they can in theory allow changes to a contract's logic and/or state. If agreement is not unanimous, the blockchain can be forked into two systems---one with the change and one without. In 2016, a significant security breach of a decentralized application called `the DAO' caused the Ethereum Foundation to propose overriding the immutability of this particular smart contract to reverse the impacts of attack. In the unusual circumstances of this case, it was possible to propose and deploy the fix before the stolen ETH could be extracted from the contract and circulated. Nodes with a philosophical objection to overriding immutability continued operating, without deploying the fix, under the name Ethereum Classic.


\subsection{Contract Migration}
\label{sec:migration}

The simplest wholesale upgrade pattern is to deploy a new version of the contract at a new address, and then inform users to use the new version---called a `social upgrade.' One example is Uniswap\footnote{\url{https://uniswap.org}}, which is on version 3 at the time of writing. Versions 1 and 2 are still operable at their original addresses. 

Contract migration does not require developers to instrument their contracts with any new logic to support upgradeability, as in many of the remaining patterns, which can ease auditability and gas costs for using the contract. However for most applications, there will be a need to transfer the data stored in the old contract to the new one. This is generally done in one of two ways. The first is to collect the state of the old contract off-chain and load it into the new contract (\eg via its constructor). If the old contract was instrumented with an ability to pause it, this can eliminate race-conditions that could otherwise be problematic during the data migration phase. The second method, specific to certain applications like tracking a user's balance of tokens, is to have the user initiate (and pay the gas) for a transfer of their balance to the new contract.
 
 

\subsection{\texttt{CREATE2}-based Metamorphosis}
\label{sec:metamorphic}

Is it possible to do contract migration, but deploy the new contract to the \textit{same} address as the original contract, effectively overwriting it? If so, developers can dispense with the need for a social upgrade (but would still need to accomplish data migration). At first glance, this should not be possible on Ethereum, however a set of opcodes can be ``abused'' to allow it: specifically, the controversial\footnote{\href{https://www.reddit.com/r/ethereum/comments/lx32kv/expectations\_for\_backwardsincompatible\_changes/}{``Expectations for backwards-incompatible changes / removal of features that may come soon.'' V. Buterin, Reddit r/ethereum, Mar 2021.}} \texttt{SELFDESTRUCT} opcode and the 2019-deployed \texttt{CREATE2}. 

Consider a contract, called Factory, that has the bytecode of another contract, A, that Factory wants to deploy at A's own address. \texttt{CREATE2}, which supplements the original opcode \texttt{CREATE}, provides the ability for Factory to do this and know in advance what address will be assigned to contract A, invariant to when and how many other contracts that Factory might deploy.  The address is a structured hash of A's ``initialization'' bytecode, parameters passed to this code, the factory contract's address, and a salt value chosen by the factory contract.\footnote{Specifically: $\mathsf{addr} \leftarrow \mathcal{H}(\mathtt{0xff} \| \mathsf{factoryAddr} \| \mathsf{salt} \| \mathcal{H} (\mathsf{initBytecode} \| \mathsf{initBytecodeParams}))$} Most often, A's initialization bytecode contains a copy of A's actual code (``runtime'' bytecode) to be stored on the EVM, and the initialization code is prepended with a simple routine to copy the runtime code from the transaction data (calldata) into memory and return. Importantly, however, the initialization bytecode might not contain A's runtime bytecode at all, as long as it is able to fetch a copy of it from some location on the blockchain and load it into memory. In order for \texttt{CREATE2} to complete, the address must be empty, which means either (1) no contract has ever been deployed there, or (2) a contract was deployed but invoked \texttt{SELFDESTRUCT}.



Assume the developer wants to deploy contract A using metamorphosis and later update it to contract B.\footnote{\href{https://medium.com/@0age/the-promise-and-the-peril-of-metamorphic-contracts-9eb8b8413c5e}{``The Promise and the Peril of Metamorphic Contracts.'' 0age, Medium, Feb 2019.}} The developer first deploys a factory contract with a function that accepts A's (runtime) bytecode as a parameter (which includes the ability to self destruct). The factory then deploys A at an arbitrary address and stores the address in a variable called codeLocation. The factory then deploys a simple `transient' contract using \texttt{CREATE2} at address T. This contract performs a callback to the factory contract, asks for factory.codeLocation, and copies the code it finds there into its own storage for its runtime bytecode and returns. As a consequence, A's bytecode is now deployed at address T. 

To upgrade to contract B, the developer calls \texttt{SELFDESTRUCT} on A. \texttt{SELFDESTRUCT} opcode wipes out the contract's code and storage of the contract account that executes the \texttt{SELFDESTRUCT} opcode. Mechanically, the consequences of \texttt{SELFDESTRUCT} on the EVM are only realized at the end of the transaction. In a followup transaction, the developer calls the factory with contract B's bytecode. The factory executes the same way placing a pointer to B in factory.codeLocation. Importantly, it generates the same address T when it invokes \texttt{CREATE2} since the `transient' contract is identical to what it was the first time---this contract does not contain contract A or B's runtime code, it just contains abstract instructions on how to load code. The result is contract B's runtime bytecode being deployed at address T where contract A was. 
 

As it is concerning that a contract's code could completely change, we note that metamorphic upgrades can be ruled out for any contract where either: it was not created with \texttt{CREATE2}, it does not implement \texttt{SELFDESTRUCT}, and/or its constructor is not able to dynamically modify its runtime bytecode.

\subsection{\texttt{CALL}-based Data Separation}
\label{sec:callbased}

To avoid migrating the stored data from an old contract to an upgraded contract, a contract could instead store all of its data in an external ``storage'' contract. In this pattern, calls are made to a ``logic'' contract which implements the function (or reverts if the function is not defined). Whenever the logic contract needs to read or write data, it will call the storage contract using setter/getter (aka accessor/mutator) functions. An upgrade consists of (1) deploying a new logic contract, (2) pausing the storage contract, (3) granting the new logic contract access to the storage contract, (4) revoking access from the old contract, and (5) unpausing the storage contract. 

An important consideration is that the layout of the storage contract cannot be changed after deployment (\eg we cannot add a new state variable). This can be side-stepped to some extent by implementing a mapping (key-value pair) for each primitive data type. For example, a new uint state variable can be a new entry in the mapping for uints. This is called the Eternal Storage pattern (ERC930). It however requires that every data type be known in advance, and is challenging to use with complex types (\eg structs and mappings themselves).

A variant of this pattern can introduce a third kind of contract, called a proxy contract, to address the social upgrade problem. In this variant, users permanently use the address of the proxy contract and always make function calls to it. The proxy contract stores a pointer (that can be updated) to the most current logic contract, and asks the logic contract to run the function using \texttt{CALL}. Unlike the functional component pattern (Section~\ref{sec:component}), the proxy will catch and forward \textit{any} function (including new functions deployed in updated logic contracts) using its fallback function.  With or without proxies, this pattern is very powerful, but instrumenting a contract to use it requires deep-seated changes to the contract code. As our measurements will show, it has fallen out of favour for the cleaner \texttt{DELEGATECALL}-based pattern (Section~\ref{sec:delegatecall}) that addresses the same issues with simpler instrumentation.

\subsection{\texttt{DELEGATECALL}-based Data Separation}
\label{sec:delegatecall}


This pattern is a variant on the idea of chaining each function call through a sequence of three contracts: proxy, logic, and storage. The first modification is reversing the sequence of the logic and storage contracts: a function call is handled by the proxy which forwards it to the storage contract (instead of the logic contract). The storage contract then forwards it to the logic contract using \texttt{DELEGATECALL} which fetches the code of the function from the logic contract but (unlike \texttt{CALL}) runs it in the context of the contract making the call---\ie the storage contract. When upgrading, a new logic contract is deployed, the proxy still points to the same storage contract, and the storage contract points to the new logic contract. Since the proxy and storage contracts interact directly and are both permanent, the functionality of both can be combined into a single contract. It is common for developers to call this the `proxy contract,' despite it being a combination of a proxy and a storage contract. 

This pattern is generally cleaner than using the previous \texttt{CALL}-based pattern because the logic contract does not need any instrumentation added to it. It is an exact copy of what the contract would look like if the upgrade pattern was not being used at all. However this does not mean the pattern in a turn-key solution. Each new logic contract needs to be programmed to respect the existing memory layout of the storage contract, which has evolved over the use of all the previous logic contracts. The logic contract also needs to be aware of any functions implemented by the storage contract itself---if the same function exists in both the storage contract and the logic contract (called a function clash), the storage function will take precedence.


The main issue with function clashes is that the proxy contract needs, at the very least, to provide an admin (or set of authorized parties) the ability to change the address of the logic contract it delegates to. This can be addressed in four main ways:

\begin{enumerate}


\item Developers are diligent that no function signature in the logic contract is equal to the signature of the upgrade function in the proxy contract (note that signatures incorporate a truncated hash of the function name, along with the parameters types, so collisions are possible). 

\item As found in the \emph{universal upgradeable proxy standard (UUPS)} (EIP-1822): implement the upgrade function in the logic contract, which will run in the context of the proxy contract. Its exact function signature must be hardcoded into the proxy contract. Every logic contract update must include it or further updates are impossible.

\item As found in the \textit{beacon proxy} pattern (EIP-1967): deploy another contract, called the beacon contract, to hold the address of the logic contract and implement the setter function for it. The proxy contract will get the logic contract address from the beacon every time it does a \texttt{DELEGATECALL}. The admin calls the beacon contract to upgrade the logic contract, while normal users call the proxy contract to use the DApp. 

\item As found in the \textit{transparent proxy} pattern (EIP-1538): inspect who is calling the proxy contract (using \texttt{msg.sender()})---if it is the admin, the proxy contract catches the function call and if it is anyone else, it is passed to the proxy's fallback function for delegation to the logic contract. 

\end{enumerate}


A drawback of the entire \texttt{DELEGATECALL}-based pattern is that logic contracts need to be aware of the storage layout of the proxy contract. In a stand-alone contract, the compiler (\eg Solidity) will allocate state variables to storage locations, and using \texttt{DELEGATECALL} does not change that, however new logic contracts need to allocate the same variables in the same order as the old contract, even if the variables are not used anymore. This can be made easier with object-oriented patterns: each new logic contract extends the old contract (inheritance-based storage). Other options include mappings for each variable type (eternal storage) or hashing into unique memory slots (unstructured storage). The \textit{Diamond Storage} pattern (EIP-2535) breaks the logic contract into smaller clusters of one or a few functions that can be updated independently, and each can request one or more storage slots in a storage space managed by the proxy contract itself.

\subsection{Evaluation Framework}
\input{sections/table}

Table~\ref{tab:eval} summarizes the pros and cons of each upgradeability pattern, omitting consensus override as it is only used in emergencies. Further detail and some take-aways from the evaluation are in Appendix~\ref{app:eval} and~\ref{app:eval2}. 


 \section{Finding Upgradeable Contracts} 
 \label{sec:proxyFinding}
 
We now design a series of measurement studies to shed light on the prevalence of the various upgrade patterns. We exclude retail changes from our measurements, because variable changes and external function calls are too commonplace to distinguish. We focus on wholesale patterns, and devote the most effort to finding contracts using the \texttt{DELEGATECALL}-based data separation pattern (Section~\ref{sec:delegatecall}) as these are the most widely used and there are various sub-types (UUPS, beacon, \etc). The other types of wholesale patterns are: 
\begin{itemize}
\item \textbf{Consensus override:} Only 1 occurrence to date (the DAO attack~\cite{dhillon2017dao}).
\item \textbf{Contract migration:} Not detectable in code; relies on social communication of the new address.
\item \textbf{\texttt{CREATE2}-based metamorphosis.} Already measured by Frowis and Bohme~\cite{frowisnot} in a broader study of all uses of \texttt{CREATE2}. They found 41 contracts between March 2019 and July 2021 that upgraded using this pattern.
\item \textbf{\texttt{CALL}-based data separation.} We conducted a quick study of 93K contracts with disclosed source code~\cite{smart_contract_sanctuary}. We identified the Eternal Storage pattern using regular expressions and found 140 instances, the newest having been deployed over 3.5 years ago. We conclude this pattern is too uncommon today to pursue a deeper bytecode-based on-chain measurement.  
\end{itemize}

\subsection{Methodology} 

\begin{figure}[t!]
  \includegraphics[width=1\textwidth]{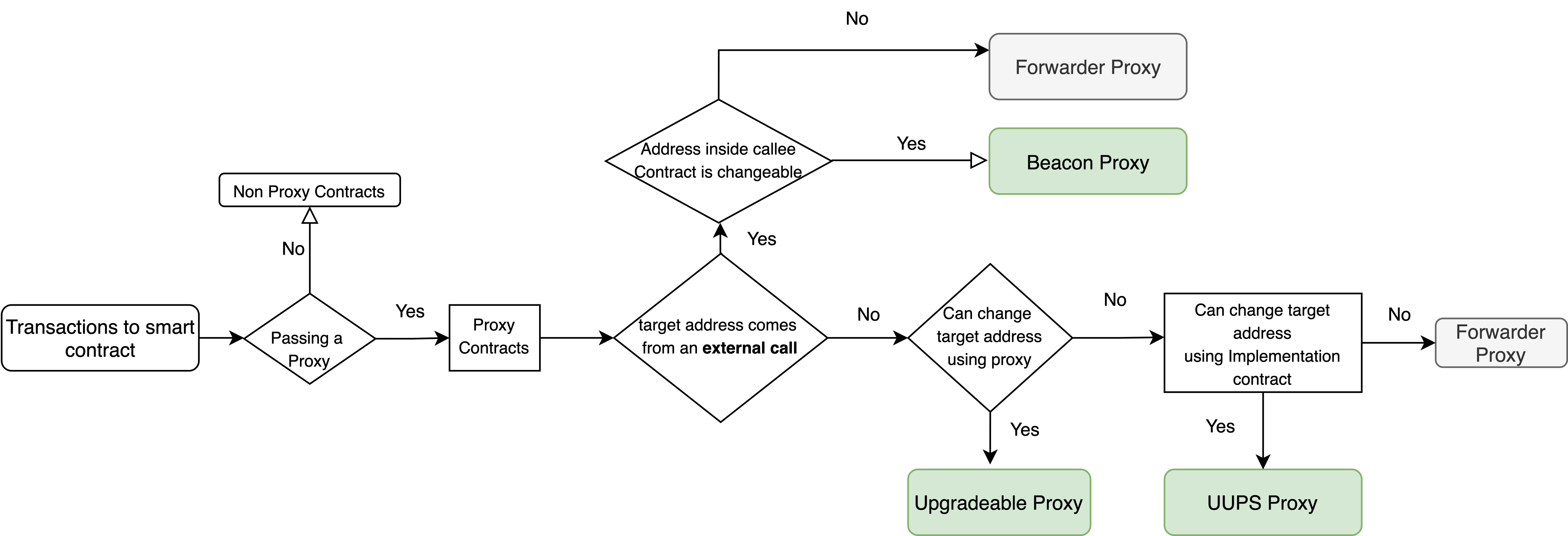}\label{flowchart}
  \caption{Flowchart for distinguishing upgradeable contracts (green) from forwarders, and for determining the upgradeability pattern type.}
\end{figure}


\paragraph{Finding proxies.} While not every use of a proxy contract is for upgradeability (\eg minimal proxies~\cite{minimalProxy}, \texttt{DELEGATECALL} forwarders~\cite{delegatecallForwarders}, \etc), all \texttt{DELEGATECALL}-based upgradeability variants have the functionality of a proxy. We therefore start by measuring the number of contracts with a proxy component, and then filter out the \textit{Forwarders} which do not enable upgradeability. To identify proxies, we examine every \texttt{DELEGATECALL} action and see if it was proceeded by a call with an identical function selector to the contract making the \texttt{DELEGATECALL} action, which indicates   the contract does not implement this function and instead caught it in its fallback function, and is now forwarding it to another contract at, what we will call, the \textit{target address}. We used an Ethereum full archival node\footnote{\url{https://archivenode.io/}} and replayed each transaction in a block to obtain Parity VM transaction traces. \texttt{DELEGATECALL} is one \texttt{callType} of an \texttt{action} within a trace. Specifically, if the data of two consecutive actions of a transaction are equal and a \texttt{DELEGATECALL} is in the second action, it shows that the transaction passes the fallback function (if any other function in the contract is called, other than fallback, then the first four bytes of the data will be changed). The \texttt{DELEGATECALL} indicates the fallback transferred the whole data to the target address without altering it, which means the contract implements a proxy.

\paragraph{Distinguishing forwarders and upgradeability patterns.} In an upgradeable contract, the target address for the \texttt{DELEGATECALL} must be modifiable. If it is fixed, we tag it as a forwarder. We define five common patterns for determining the target address cannot be changed:

\begin{enumerate}
  \item The target address is hardcoded in the contract.
  \item The target address is saved in a constant variable type.
  \item The target address is saved in an immutable variable type and the deployer sets it in a constructor function.
  \item The target address is defined as an unchangeable storage variable.
  \item The proxy contract grabs the target address by calling another contract but there is no way the callee contract can change this address.
\end{enumerate}

In the first three situations, the target address will be appeared in the runtime bytecode of the contract. For every proxy-based \texttt{DELEGATECALL}, we obtain the target address from the transaction's \texttt{to address}, and we obtain the caller's bytecode by invoking \texttt{eth\_getCode} on the full node. If we find the target address in the bytecode, we mark it as a forwarder. 

In the fourth case, we find where the target address is stored by the contract by decompiling the contract, with \textit{Panoramix}\footnote{\url{https://github.com/palkeo/panoramix}}, locating the line of code in the fallback function that makes the \texttt{DELEGATECALL}, and marking the storage slot for the target address. We parse the code and check if an assignment to that slot happens in any function in the contract---this is non-trivial and we refer the interested reader to Appendix~\ref{app:assignment} for the full details. If any assignment is found, we should be sure that the other variable assigned to the target address variable comes from the input of that function. If these conditions are satisfied, there is a function inside the contract that can change the target address and we mark the proxy as an upgradeable proxy contract. 

Recall in the Universal Upgradeable Proxy Standard (UUPS) pattern, the logic contract implements a function to update the target address that is run in the proxy contract's context using \texttt{DELEGATECALL}. This is a subcase of the fourth case, where we check the logic contract instead of the proxy contract. If we determine the logic contract can assign values to the logic contract in any function, we tag it as UUPS.

In the fifth case, we rewind the transaction trace from the proxy-based \texttt{DELEGATECALL} and look for the target address being returned to the proxy contract in another action. If we find it being returned by a contract, we apply the methodology from the fourth case to this contract. If the target address is modifiable, we mark it as using the Beacon proxy upgradeability pattern. All contracts that remain after performing all of the checks above are marked as forwarders. 

\subsection{Results}

\begin{table}[t]
\centering
\begin{tabular}{|l|r|}
\hline
Proxy Contracts (Total) & 1,427,215  \\ \hline 
Proxy Contracts (Filtered) & 13,088  \\ \hline 
Regular Upgradeable Contracts & 7,470  \\ \hline
UUPS & 403  \\ \hline
Beacon & 352  \\ \hline
\end{tabular}
\caption{\label{tab:updata} Results of each \texttt{DELEGATECALL}-based upgrade pattern for the time-period {Sep-05-2020} to {Jul-20-2021} (2,064,595 blocks).}
\vspace{-10pt}
\end{table}

Our measurements cover block number \texttt{10800000} to \texttt{12864595}, which corresponds to the time-period \texttt{Sep-05-2020} to \texttt{Jul-20-2021}, and are reported in Table~\ref{tab:updata}. While we found 1.4M unique proxy contracts, many of these share a common implementation contract and are part of the same larger upgradable system. As one example, the NFT marketplace OpenSea~\footnote{\url{https://opensea.io}} gives each user a unique proxy contract. After clustering contracts, we find 13K unique systems.  

For the 8,225 upgradeable systems (regular, UUPS and beacon), we randomly sampled 150 contracts and manually verified they were upgradeable proxy contracts. We also sampled 150 contracts from the forwarders to verify they are not upgradeable, however we did find 2 false-negatives. Our model did not catch these contracts because a failure happened when decompiling them and our assignment checker detector in turn failed. Note that for UUPS contracts, the implementation contracts are much larger and harder to analyze than the proxy contract itself.


\section{Finding the Admin}
\label{sec:governance}

If a contract is upgradeable, someone must be permissioned to conduct upgrades. We call this agent the admin of the contract. In the simplest case, the admin is a single Ethereum account controlled by a private signing key, called an externally owned account (EOA). A breach of this key could lead to malicious updates, as in the case of the lending and yield farming DeFi service Bent Finance ~\cite{bentFinanceHack}. Bent Finance deployed a \textit{Transparent Upgradeable Proxy} with an EOA admin that was breached (unconfirmed if via an external hack or insider attack). The EOA pushed an updated logic contract\footnote{\url{https://etherscan.io/address/0xb45d6c0897721bb6ffa9451c2c80f99b24b573b9}}  which moved tokens valued at \$12M USD into the attacker's account\footnote{0xd23cfffa066f81c7640e3f0dc8bb2958f7686d1f} and then upgraded the logic contract to a clean version to cover-up the attack. Based on \textit{The State of DeFi Security 2021}~\cite{certikReport} report by Certik,\footnote{\url{certik.com}} ``centralization risk'' is the most common attack vector for hacks of DeFi projects. 
 


Control over upgradeability typically falls into one of three categories: 


\begin{enumerate}
\item \textbf{Externally owned Address (EOA):}
One private key controls upgrades. It is highly centralized and one malicious admin or compromised private key could be catastrophic. It is also the fastest way to respond to incidents. An EOA may also pledge to delegate their actions to an off-chain consensus taken on any platform, such as verified users on \textit{Discord} or \textit{Snapshot}, however with no guarantee they will abide by it. In our measurements, we cannot distinguish this subtype as these are off-chain, social arrangements. 

\item \textbf{Multi-Signature Wallet:}
Admin privileges are assigned to a multi-signature wallet, requiring transactions signed by at least $m$ of a pre-specified $n$ EOAs.   This distributes trust, and tolerates some corruption of EOAs or loss of keys. There is no guarantee different EOAs are operated by different entities and may be security theatre put on by a single controlling entity.


\item \textbf{On-Chain Governance Voting:} A system issues a governance token and circulates it amongst its stakeholders. Updates are decided through a decentralized voting scheme where the weight of the vote from an EOA (or contract address) is proportionate to how many tokens it owns. This system is potentially highly decentralized, but the degree depends on the distribution of tokens (\eg if a single entity controls a majority of tokens, it is effectively centralized). Voting introduces friction: (1) a time delay to every decision---some critical functionality might bypass the vote and use quicker mechanisms (\eg global shutdown in MakerDAO), and (2) on-chain network fees for each vote cast.

\end{enumerate}

\subsection{Methodology}


We conduct our measurement on the 7,470 regular upgradeable contracts from Section~\ref{sec:proxyFinding}. The process can be divided into two main parts: finding the admin account's address and finding the admin type (EOA, multi-sig, or decentralized governance).

\paragraph{Finding the admin account's address.} EIP-1967 suggests specific arbitrary slots for upgradeable proxy contracts to store the \textit{admin address}.\footnote{Storage slot 0xb53127684a568b3173ae13b9f8a6016e243e63b6e8ee1178d6a717850b5d6103} We first check this specific storage slot using \texttt{eth\_getStorageAt} on the full node. If it is non-zero, we mark what is stored as the admin address. 
For non-EIP-1967 proxies, we use a process that is very similar to how we found the storage slot of the target address in Section~\ref{sec:proxyFinding}. We first find the function in which the admin can change the \textit{target address} (upgrade function). This function is critical and should only be called by the admin. We extract the access control check and mark the address authorized to run this function as the admin address.

\paragraph{Finding the admin type.} Having the admin address, we can check if the account is an EOA by invoking \texttt{eth\_getCode} on the address from the full node: if it is empty, it is an EOA. Otherwise, it is a contract address. The most common multisig contract is Gnosis Safe.\footnote{\url{https://gnosis-safe.io/}} We automatically mark the admin type as multi-sig if we detect Gnosis safe. We then switch the manual inspection to find other multi-signature wallets (\eg MultiSignatureWalletWithDailyLimit, \etc) and add them to the data set. 

In some cases, the admin address is itself a proxy contract---a pattern known as an Admin Proxy. This adds another layer of indirection. We are reusing our methodology for identifying proxy contracts to exact the real admin account, and the proceed as above. Further details of the methodology and implementation are provided in Appendix~\ref{app:admin}.

\subsection{Results}

\begin{table}[t]
\centering
\begin{tabular}{|l|r|r|r|r|r|r|}
\hline      &\multicolumn{2}{c|}{EIP-1967} & \multicolumn{4}{c|}{Non-EIP-1967}         \\
\hline
Type 				& \makecell{Regular\\Admins} & \makecell{Admin\\Proxy} 	& \makecell{Regular\\Admins} 	& \makecell{Admin\\Proxy} & \makecell{Arbitrary\\Slots} &  \makecell{Fixed\\Address}  \\ \hline 
EOA 				& 900  	& 1202		& 1313		& 92			& 2		& 49 		\\ 
Multisig 			& 255  	& 567		& 104		& 16			& 10		& 36		\\ 
Governance/Other 	& 53  	& 462		& 			& 			& 		& 160 	\\ \hline
\end{tabular}
\caption{\label{tab:admindata} Results of each admin type in upgradeable contracts for the time-period {Sep-05-2020} to {Jul-20-2021} (2,064,595 blocks).}
\vspace{-10pt}
\end{table}

Of 7470 proxies, 3558 are controlled by an EOA address, 988 are controlled by a known multi-signature wallet, and 2924 addresses are remaining. Table~\ref{tab:admindata} breaks down each sub-category for these. Of the latter 2924 addresses, these are either decentralized governance or another unknown type. After manual inspection, we note some of the unknown contracts use undefined or new patterns for implementing multi-sig contracts; our model has false negatives in detecting multi-signatures. The results demonstrate significant centralization risk in upgradeability: 48\% of systems could be upgraded with the breach of a single signing key, and an additional 13\% by potentially a small number of signing keys.




\section{Concluding Remarks}


In our paper, we find that \texttt{DELEGATECALL}-based data separation is the most prominent upgrade pattern in Ethereum in recent years. Our evaluation framework gives some hint as to why this is the case. It avoids the need for a social upgrade, as in contract migration or the \texttt{CALL}-based pattern (without a proxy). \texttt{CREATE2}-based metamorphosis was recently made possible (with the introduction of \texttt{CREATE2}) and its use might grow over time, however it shares one major drawback with contract migration: the need to migrate the whole state from the old contract for each update, even if the update makes minor changes to the logic of the contract. Metamorphic contracts also run the risk of Ethereum removing the \texttt{SELFDESTRUCT} opcode they rely on.  A drawback of \texttt{CALL}-based patterns is the heavy instrumentation each new contract needs before it can be deployed, whereas in a \texttt{DELEGATECALL}-based (along with migration and \texttt{CREATE2}-based) upgrade pattern, developers can simply deploy the new logic contract exactly as it is written. Putting these reasons together, \texttt{DELEGATECALL}-based pattern is an attractive option on balance. 


The main take-away from studying upgradeability on Ethereum is that immutability, as a core property of blockchain, is oversold. Immutability has already been criticized for being dependent on consensus---both technical and social~\cite{walch2016path}---however the widespread use of upgradeability patterns further degrades immutability. Finally, as we show, the prominence of contracts that can be upgraded with a single private key (\ie externally owned account) calls into question how decentralized our DApps (decentralized applications) really are. If the upgrade process is corrupted through a key theft or by a rogue insider, the whole logic of the contract can be changed to the attacker's benefit. 

One recent application of our research was finding all contracts that implement the UUPS upgrade pattern, which become important when a vulnerability is discovered in one of the best-known libraries for implementing UUPS. We describe how we can find potentially vulnerable contracts in Appendix~\ref{sec:attackUUPS}. While others had found some contracts by looking for specific artefacts left by the UUPS library, we improved the state of the art by looking for the generic pattern of UUPS. 

A final discussion point concerns Layer 2 (L2) solutions, such as optimistic rollups and zk-rollups~\cite{mccorry2021sok}. For the readers that are already familiar with them, their central component is a bridge contract that let computations be performed off of Ethereum (layer 1) and have just the outputs validated on Ethereum. If the bridge contracts is upgradeable, the rules for accepting L2 state are also upgradeable which means every L2 contract is de facto upgradeable even if it does not implement an upgrade pattern. We saw Ethereum override the consensus of the network to revert the DAO hack, which was a rare and contentious event. If a similar attack happened on a L2, reverting would be much simpler and not require a hard fork: the L2 could simply update the bridge contract. For this reason, the consensus override upgrade pattern may be less rare in the future.


%% file: sections/table.tex

\begin{table*}[ht!]

    \renewcommand{\arraystretch}{1}
    
    \centering
    
    \begin{tabular*}{0.9\textwidth}{@{\extracolsep{\fill}} lcccccccccccc}
    
    \textit{Method} &
    \headrow{Can replace entire logic} & 
    \headrow{No need to migrate state from old contract} &  
    \headrow{User endpoint address unchanged} &
    \headrow{No need to instrument source} &
    \headrow{No need to deploy a new contract to upgrade} &
    \headrow{No indirection between contracts} & 
    \headrow{No downtime to upgrade} &
    \headrow{No function selector clashes} &
    \headrow{No storage clashes} & 
    \headrow{ } & 
    \headrow{ } \\ \hline 
    
    Parameter Configuration 	&	  &\full&\full&\full&\full &\full&\full&\full&\full&&\\
    Component Change 	        &	  &\full&\full&\prt	&      &\prt &\full&\full&\full&&\\
    Contract Migration			&\full&     &     &\full&	   &\full&\full&\full&\full&&\\ 
    Create2 metamorphosis 		&\full&     &\full&\full&      &\full&	   &\full&\full&&\\
    Call-based 		            &\full&\full&     &     &	   &	 &\full&\full&\full&&\\ 
    DelegateCall-based 			&\full&\full&\full&\prt	&	   &     &\full&     &     &&\\ \hline 
    \\
                                                                                        
    \end{tabular*}
    
    \caption{An evaluation of upgradeability patterns. \full~ indicates the upgrade method is awarded the benefit in the corresponding column. \prt~partially awards the benefit. Empty cells shows that the method does not satisfy the property.} 
    \label{tab:eval}
    \end{table*}

%% file: sections/appendix.tex


 \section{Evaluation Framework: Details}
 \label{app:eval}
 
 In this section we compare and evaluate different methods discussed in previous section and explain the consequences regarding each method to the users and developers of Dapps.
 \subsection{Criteria}
 There are some characteristics that can help the designer to decide which method should be used on the system and add upgradeability to the Dapp. In this part we pencil out these criteria and evaluate different methods based on these criteria. In this part we describe and specify what it means that each row of our table receives a full dot (\full), partial dot (\prt), or nothing. 
 
\subsubsection{Can replace entire logic}
An upgradeability method in which the admin is able to replace the entire logic of the system earns a full dot (\full) otherwise it receives nothing.

\subsubsection{No need to migrate state from old contract}
In some patterns, there is no need to collect data from the old version and push it to the new contract which receive a full dot (\full). On the other hand, patterns which required to migrate data from old version receive nothing.

\subsubsection{User endpoint address not changed}
In some upgradeability methods, after the upgrade process, users must call a new contract address to use the Dapp. It is equivalent to having 2 different Dapps at the end of the upgrade. Alice uses Dapp X which uses one of the upgradeability patterns at address A before the upgrade. After upgrade, she may be unaware that upgrade happened and use the previous address (receive full mark (\full)) or she may need to use address B instead which receive nothing.

\subsubsection{No need to instrument source}
Upgradeability patterns in which the developers do not need to change any part of the original code to add the upgrade method receives nothing. The methods in which the developers do not need to change the whole code but should add a proxy contract or change just one component of the system receive half dot (\prt) and patterns in which the developers should change the whole code to add upgradeability receive full mark (\full).

\subsubsection{No need to deploy a new contract}
In some upgradeability patterns, the admin needs to deploy a new smart contract in the process of upgrade which receives nothing. Upgradeability methods which do not need to deploy a new contract for the process of upgrade receive a full dot (\full).

\subsubsection{No indirection between contracts}  
Indirection happens if the first external message need be forwarded from a contract to another using one of the \texttt{CALL}, \texttt{STATICCALL}, or \texttt{DELEGATECALL} opcodes. Upgradeability methods that do not need any indirections receive a full dot (\full). An upgradeability pattern that contains indirections which adds an extra gas because of adding one or more layers of indirection awarded nothing. An upgradeability method in which not all but just a portion of the incoming transactions need indirection receive half dot (\prt).

\subsubsection{No downtime to upgrade}
Patterns which have a downtime of the Dapp in the upgrade event receive full dot (\full) otherwise it receives nothing. 

\subsubsection{Function Selector Clashes}
Upgradeability methods in which the developer should take care of function selector risks due to the using of \texttt{DELEGATECALL} opcode receive full dot (\full), otherwise receive nothing. 

\subsubsection{Storage Clashes}
Upgradeability methods in which the developer should take care of storage clashes risks in two contracts due to the using of \texttt{DELEGATECALL} opcode receive full dot (\full).


\clearpage
 \section{Evaluation Framework: Take-Aways}
 \label{app:eval2}
 
In this section we discuss about the consequence of each upgrade methods regarding the criteria we mentioned in the previous part for users and developers that want to use the upgradeability pattern or uses a Dapp that uses one of the mentioned patterns.

\subsection{Speed of an Upgrade}
Upgrade events of a Dapp consists of two different processes. First a way to come to an agreement for changing the system, and then a way to implement and execute the change. The First part depends on the reason behind the upgrade. If the upgrade is to patch a bug, then the process to come into agreement is very fast but if the goal behind upgrade is to add new functionality or change a logic, it usually starts with a proposal and after some discussions, if the agent that responsible for the decision agree with the proposal, the execution part will be started. We won't discuss the first process because it depends on the type of agent discussed in \ref{sec:governance}.
After coming into agreement about the change, the speed that the admin can implement and execute the upgrade depends on two three criteria discussed above: \textit{No need to deploy a new contract to upgrade}, \textit{No need to migrate state from old contract}, and \textit{having a downtime in the upgrade process}.
 
\textit{Parameter change} method is the fastest way to execute the upgrade because there is no need to deploy a new contract, and no need to migrate state and no downtime in the system.
\textit{Component change} method change is not as fast as Parameter change method but faster than other types because the admin needs to deploy a specific smart contract which is a small component of the system and also update an address variable inside the main contract that points to that specific component and change it to the address of the new version of that component. But there is no need to migrate data and there is no downtime needed for this upgrade method.

\textit{Migration} method has a slow upgrade process. The reason is that the admin needs to deploy a new contract and also the admin or users should transfer the data from old version to the newer version. In most migration processes the developer team deploy a \textit{Migrator} contract and users should use this Migrator contract to withdraw their funds/data from the previous version and move it to the newer version. But, there is no downtime in the Dapp and no need to change a state variable.

\textit{Call-based} and \textit{DelegateCall-based} are very similar to each other in the speed of upgrade. These two are not as quick as \textit{Retail changes} because the developer needs to implement and deploy the \textit{whole logic} contract to the blockchain and then change the pointer addresses inside the storage/proxy contract to the newer version.
On the other hand these two approaches are faster than \textit{Migration} because as mentioned before, there is no need to migrate data. There is no downtime in these methods.

\textit{Metamorphic} method is the slowest way to upgrade a system which uses this method because similar to the migration plan there is a need to deploy a contract and migrate the state to the newer version but there is a difference between these two. In metamorphic method, the admin first should \textit{Self-Destruct} the previous version in a single transaction and after that transaction send a contract creation transaction to deploy the newer version. Because self-destruct happened at the end of the transaction, the process of upgrade happens on two different transactions which is a downtime to the system. This downtime could be a gap between order of the two transaction in a single block or could be gap between blocks that these two transactions included into blockchain.

\subsection{Cost of Upgrade}

 One of the main differences between upgradeability approaches is how much does the upgrade process costs for the admins and users. The cost of upgrade mostly depends on two criteria explained above: 1) no need to deploy a new contract, 2) no need to migrate a the state to newer version.

 \textit{Parameter change} method is the cheapest method in the upgrade event because there is no need to deploy a new contract or migrate data.
 \textit{Component change} is in the middle, because there is a need to deploy a new contract (however it is cheaper comparing to methods in which we should deploy the whole logic), but there is no need for data migration.
 \textit{Migration} plan is very expensive in the upgrade event because we need to deploy a new contract and migrate the data from the old version which is very expensive.
 \textit{Call-based} and \textit{DelegateCall-based} are very similar to each other in the cost of upgrade which is more expensive than component change, but cheaper than migration. In both the admin must deploy the a contract containing the whole logic, but there is no need to migrate the whole data.
 \textit{Metamorphic} method is the most expensive method we have because we need to deploy a new contract, migrate data to the newer version and also we need to self-destruct the previous version before the upgrade event which adds cost to the upgrade process.

\subsection{Gas overhead for users}
Sometimes in upgradeability patterns, we have a tradeoff between adding a feature to the pattern to improve it and increasing the cost (gas needed for the transactions) for users that want to interact with our Dapp.

In patterns that needs indirection, such as \textit{Call-based}, \textit{Delegatecall-based}, and \textit{Component change} pattern an extra cost will be added to the users, because for all or some of the transactions to the Dapp, our system needs to forward the calls to another contract using Call or Delegatecall opcode to the users. 
Also in \textit{Delegatecall-based} pattern to mitigate the function selector clashes or storage clashes, we need to add some other checks to our code which also increases the cost of interacting with the Dapp.
Also there are some other ideas that addresses some limitations of one type of upgradeability pattern, but increases the cost for users. For instance, in \textit{Call-based} approach one of the problems is that after upgrade users should use a new address for using the Dapp but adding a \emph{Registry} contract can help to mitigate this. Using Registry contract, all other contracts should ask the registry to find out the latest version of the contract and then calls to the newer version which adds a gas cost to the users. 
 
 \subsection{Useability}
 Upgradeability patterns differ in term of Useability and it depends on three criteria explained above; \textit{User endpoint address unchanged}, \textit{No need to migrate state from old contract} and \textit{Downtime in upgrade events}.

Patterns in which the endpoint address is changing after upgrade event, \textit{Migration} and \textit{Call-based} is not user friendly because each time that the upgrade happens, the user must use the newer address. So there is need to make awareness about this address change which is a hard action and need to socially interact with the all users and make them aware of the change. We have two main type of users in the Dapp ecosystem, normal user or another smart contract (Dapp) that uses our system. Regular users which uses the official interface (website) of the project may do not sense any changes, but users that work with the smart contract directly or via their own interface (e.g, Centralized Exchanges), or other Dapps that uses the smart contract must have a way to upgrade the address they uses to use the newer version and if they did not implement a way to upgrade this address then their Dapp will face problems. So these patterns are make problems for composablility of the ecosystem.

In most of \textit{Migration} plan upgrade events, users are responsible for the migration of their data using a migrator contract (for instance, the user must withdraw the fund and use a migrator contract to push the data into the newer version) which add costs to the user and it is not user friendly. This is one reason that make the migration plans very hard because some users are not doing the process of migration and stay on the previous version which is like having a fork for the Dapp in side of the Dapp team (e.g, Uniswap V2 and V3).
In \textit{Metamorphic} pattern as mentioned before there is a downtime during the upgrade. So users cannot work with the Dapp on that exact time which is not user friendly.

\subsection{Dealing with two different new versions}
In \textit{Migration} and \textit{Call-based} pattern we will come up with two different Dapps after each upgrade event. So, a decision must be made for the previous version. One possible choice could be shutting down the old version. It can be done by self-destructing the old version, or by having a pausing mechanism to stop the older version functionality. In migration plan it is not regular to stop the previous version because in most migration plans, users are responsible to move their funds and data from the previous version to the new one and we cannot force them to do that, so we cannot stop the smart contract. 
The other option could be having a mechanism that after the upgrade, all calls to the previous version just be forwarded to the newer version which add costs and have some limitations like we cannot call the new functions defined in the newer version using the old version. This option is doable in Call based patterns. The other problem of this option is that if we upgrade a system more than one time then the calls to the first version should be redirected through lots of contracts to reach to the newer version. Also it adds complexity because developers must maintain more than one contract~\cite{tobBlogPost}.

 \subsection{System Complexity} \label{sysComplexity}

 Using upgradeability patterns will add to complexity of our system but the degree of complexity varies and depends on the pattern. 
\textit{Parameter Change} method does not change the system in general but just adding a mechanism to change pre-specified variables in the system. The most important issue about this pattern is that the developer team must limit the boundary of these parameter for the security of the system. For instance in MakerDao platform~\footnote{\url{https://makerdao.com/}}, Stability fee is changeable but if this variable be changed to 100\% then the whole system will be halted, so it should be limited.
\textit{Component Change} pattern is very similar to the parameter change, but a whole component could be changed and finding the safe boundary of changes and limiting this boundary is a bit harder.
\textit{Migration} plans for upgradeability does not change any complexity to the system because we do not need to change any part of system to add this type of upgradeability to it. The only important issue regarding this pattern is that we must be sure that there is a way to collect data from the old version like having getter functions for reading data and also having a withdraw function for users to collect data and funds from previous version and push or deposit it to the newer version.
Using \textit{Call-based} patterns adds higher degree of complexity to the system compared to previous patterns. As discussed before, in this pattern we must be sure that the storage and logic contract is divided and there is not any storage variable inside the logic contract. This is one of the main security issues that found in the Dapps using this pattern regarding Trail of Bits company reports~\cite{tobBlogPost}. To add a way in storage contract to define new variables, developers uses the eternal Storage pattern for their storage contract which is very hard to apply for complex data structures in Ethereum such as mappings or structures. This is another source of complexity using Call-based pattern.
\textit{Delegate-call} pattern adds complexity to the code because of using \textit{Delegate-call} opcode in its logic. As mentioned above because of using this opcode, the developer should take care of storage clashes and also function selector clashes. Other than these two there are some other limitations and risks of using this patterns. For instance, we cannot have a \textit{Constructor} function on the logic contract, because constructor functions is used to initialize specific variables at deployment time and if we have a constructor inside the logic, then storage of implementation contract will be changed and not storage of proxy contract. To mitigate this problem we can add a regular function named \textit{Initialize} function inside the implementation and make sure that this function can be called \emph{once} to act just like a constructor function. 
\textit{Metamorphic} pattern is proposed recently and not well-tested yet. There are some risks to this pattern as well. We should be sure that we have a mechanism to self-destruct the contract. Otherwise we cannot redeploy a new version and so our contract won't be upgradeable. The other important issue related to Metamorphic pattern is that the developer must know that each time they want to upgrade the system the whole storage will be wiped out and need to re-initiate the whole state after re-deployment.

\clearpage 

\section{Assignment Checker Module}
\label{app:assignment}

\begin{figure}[t!]
  \centering
  \includegraphics[keepaspectratio,height=22cm, width=13cm]{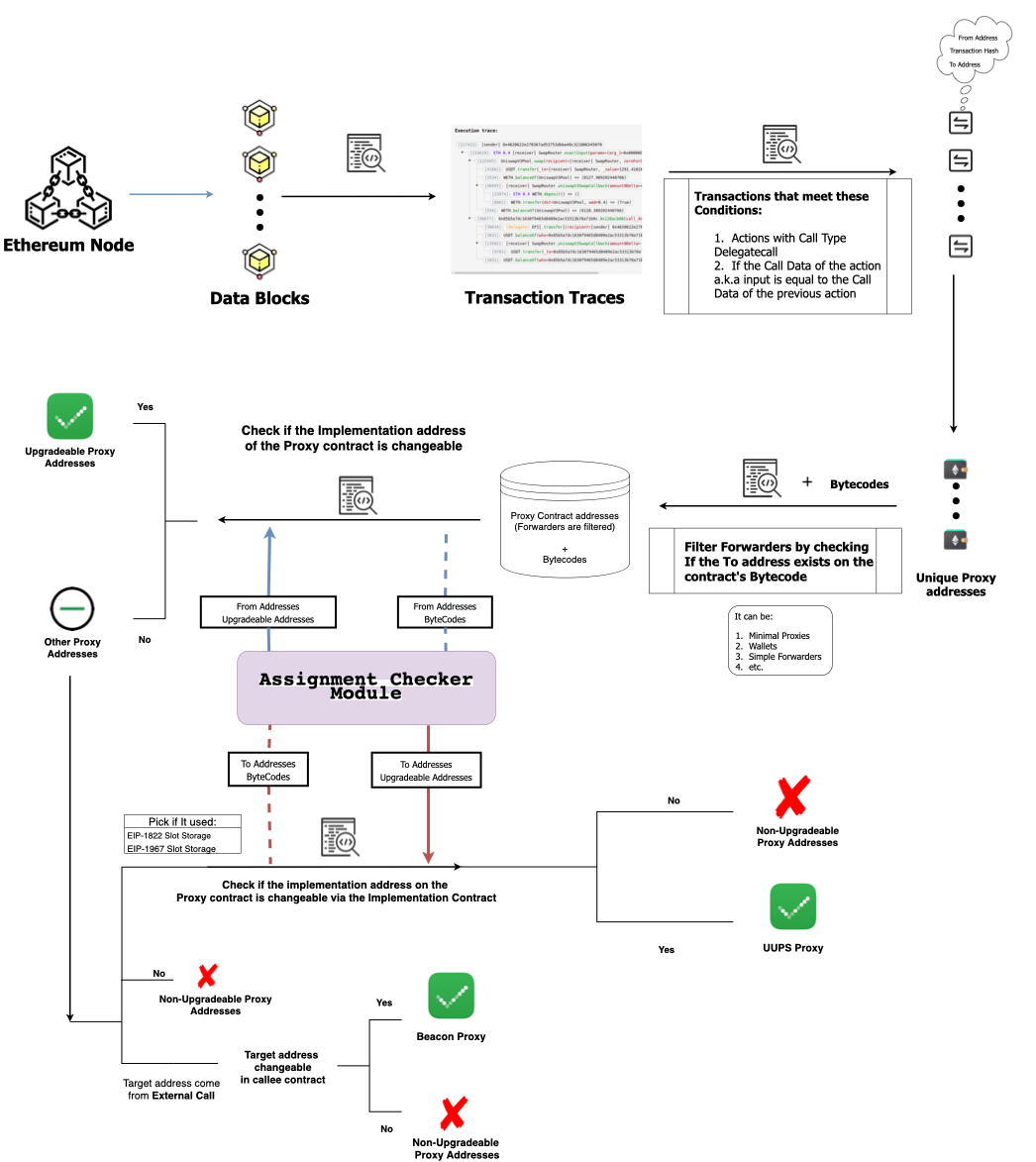}\label{fig:finderModule}
  \caption{Upgradeability Proxy Contract Finder}
\end{figure}

The whole measurement process is depicted on figure~\ref{fig:finderModule}. We need a module to check whether the admin can change target address on the proxy contract,using a function in the proxy contract, implementation contract or beacon contract. For this purpose the module must get the \textit{Bytecode} of the proxy,implementation or beacon address as input and find the variable name and also its storage slot of the target address. Then checks to find out is there any function inside the contract that gives the admin the ability to change the target address.

We use bytecode decompiler named \textit{Panoramix decompiler} \footnote{\url{https://github.com/palkeo/panoramix}} to decompile the bytecode into well-formatted python language codes. The decompiled code gives us all storage variables of the related contract and the storage slots of those variables in a function named \textbf{Storage}. On the other hand, the decompiled code will tell us if a function is \textit{Payable} or not. Among these Payable functions the one that does not have name or its name is fallback is the \textit{fallback} function of the contract. So we will try to find the line of code that \textit{Delegate Call} happened on it and collect these lines. Now that we have storage variable names and storage slots of these variables and also the line of code inside fallback that have the delegatecall, we will check to find the target address variables. We are doing that by checking if one of the storage variables inside Storage function is used in the line of code that contains delegate call. We will add them to an array of implementation addresses.

There is two other steps here. First finding other variable names with the same storage slots as the implementation addresses we found from the first step by checking the Storage function and also finding another variables that being assigned to those implementation variables in some other part of the code. We will add these two type of variables to the implementation addresses as well.

Now that we have a list for implementation addresses, we will search through the code to find if any assignment happened to one of them. If yes we will pick the variables that is assigned to target variable and then check if this assignment happened in a specific function and to one of the inputs of that function. In this case this function will be the upgrade function because the caller of this function can upgrade the target address by calling this function with desired input. 

To summarize what we did, we find all possible variables in the code that can change the target address inside the contract and check if there is any function inside them that can assign new address to the target address variable.

\begin{figure}[t]
  \centering
  \includegraphics[keepaspectratio,height=13cm, width=12cm]{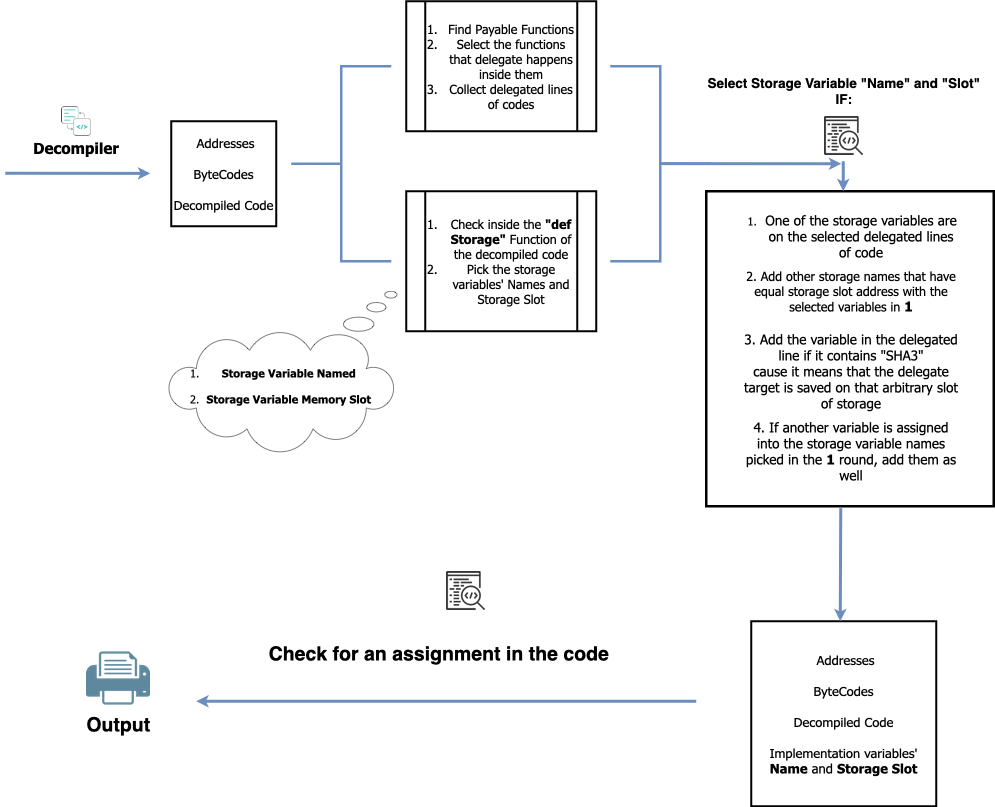}\label{assignmentFinder}
  \caption{Assignment CheckerModule}
\end{figure}

The whole process is depicted on figure~\ref{assignmentFinder}.

\clearpage
\section{Detailed Methodology for Finding Admin Types}
\label{app:admin}

In this section, we explain the way we find admins of the proxy contracts described in Section~\ref{sec:governance} in more details regarding the methodology and results.

\paragraph{EIP-1967.}As mentioned above EIP-1967~\footnote{\url{https://eips.ethereum.org/EIPS/eip-1967}} suggested specific arbitrary slots for upgradeable proxy contracts to store implementation contract's address and \textit{Admin address}\footnote{Storage slot 0xb53127684a568b3173ae13b9f8a6016e243e63b6e8ee1178d6a717850b5d6103 for admin}.

In first step we use \textit{eth\_getStorageAt} method of an Ethereum full archival node to search the EIP-1967 specified storage slot for admins on our 7,470 proxy contracts. If the result of this method is non-zero it means that the proxy uses EIP-1967 standard because the specified storage slot is an arbitrary slot and one can store variable in this slot just by defining this slot which means that they used EIP-1967.

So, for non zero results, we capture the address which is the address of admin of the proxy. Now we try to find the type of these admin addresses. Having the address of the admin we use \textit{eth\_getCode} method to check the code of the admin account. If the code is empty, it means that this account is not a smart contract so it is an EOA. we find \textit{900} EOA admins that their proxy uses EIP-1967 standard.

The remained admin addresses are contract because their account keeps code. This contract can be multi signature smart contract wallets. The most widely used multi signature wallet is Gnosis Safe\footnote{\url{https://gnosis-safe.io/}} wallets. We automatically checked if the code of the admin address is the Gnosis wallet multi signature patterns. After picking Gnosis safe wallets we manually checked \%10 of the remained addresses to find if they used other patterns for their multi signature wallet and we found some other patterns (e.g. MultiSignatureWalletWithDailyLimit, etc.). After Finding all these types we checked the admin codes to see whether they are multi signature wallets. We find \textit{255} admin accounts that uses multi signature wallets as their admin.

There is another class of admin contracts named \textit{Admin Proxy} contracts. These admin proxy contracts are another layer of re-direction between the real admin and the Dapp's proxy contract. The admin proxy contracts are proxy contracts that redirect the messages from the real admin into the Dapp's proxy. The only person who can use admin proxy is the admin (a.k.a owner) of the admin proxy. So we first filter the admin proxy contracts using the codes we get from the previous part and then try to find the owner of the admin proxy contracts. The owner of admin proxy contract (the real admin) also can be EOA, Multi-sig or governance contract. Finding the owner of the admin proxy contract, we used \textit{eth\_getCode} method to check the code of these account and find out if they are EOAs or Multi-signatures or governance schemes. Doing this we find \textit{1202} EOA admin accounts and \textit{567} multi signature admins. We marked the remained proxy admin addresses as Governance/Not Known admin types and we have \textit{462} of them. There were also non admin proxy contracts which use EIP-1967 but they were not EOA or Multi signatures. We marked them as Governance/Not Known admin types and we have \textit{53} of them.

\paragraph{Non EIP-1967.}

For proxy contracts which not use EIP-1967, the problem is we don't know where the admin address is saved in the proxy contract's storage (what is the storage slot of the admin address). It can be saved in a storage slot of the contract or be hardcoded in the smart contract\footnote{There are some other possible ways to store the admin address for instance saving it in another contract and each time make an external call to get the address but to our knowledge this pattern is not widely used as a standard}. 

So there are two ways that the admin address is saved in the proxy contract. It can be saved as a storage variables or it can be hardcoded as a fixed address.

In storage variable case, the first question is in which storage slot the admin address is stored. So, the first step is to find the storage slot of the admin address variable. Also for the fixed address we should find the fixed address of the admin directly.

To find the slot of the storage variable in which admin address is saved, we first find the function in which the proxy can be upgraded. For finding the upgrade function we exactly do what we did in \ref{sec:proxyFinding} part. We first find the storage variable in which we saved the implementation address and then we find a function in which the implementation address can be changed using the inputs of that specific function. 

The upgrade function of a proxy contract is a critical function and the only account that can call this function should be the admin of the proxy contract. So, there should be an access control check inside the upgrade function to check whether transaction sender is equal to the admin address or not.
So, after finding the upgrade function we search for conditionals that checks the caller of the transaction and by doing that we can find the admin address or the storage variable in which the admin address is stored.

If the admin address is stored in a storage variable, then we should find the storage slot of that specific storage variable. For finding the storage slot we do what we did in \ref{sec:proxyFinding} part by using \textit{def storage} function of the decompiler and check the storage slot of the storage variable we found, and the admin address is saved on it. 
Now we have the storage slot of the admin address and we should start doing all the things we did for EIP-1967 in the previous part. 
In the EIP-1967 the storage slot for admin address was pre-specified and we do not need to find the slot but in this case we use the above methodology to find the slot but the further steps are the same as EIP-1967.
So, by using the \textit{eth\_getCode} method for admin address inside the storage slot we find above, we can check wether the admin is EOA, Multi-sig, Governance, Proxy admin or not known.
In this part we find \textit{1313} EOA addresses and \textit{104} multi-sig admins. Also by checking proxy admins we find \textit{92} EOA addresses and \textit{16} Multi signatures that uses proxy admin as a level of indirection. 

In another case the admin address may be stored directly in a specific arbitrary storage slots. In this type the compiler will specify the address using the \textit{sha3} hash notation. In this case same as above we find the conditional check on the transaction sender and then find the storage slot in that line and hash of that pre-specified string. 
By finding this arbitrary storage slot and doing the same processes we did in the previous part we find \textit{2} EOA addresses and \textit{10} Multi-sig addresses.

The only case that is left is proxy contracts, in which the address of the admin is hardcoded inside them. It very straight forward. We find the upgrade function and the access control check on the caller of the transaction and then pick the fixed admin address and do the same processes mentioned above to find the admin types. There are \textit{49} EOAs, \textit{36} multi-signature admins and \textit{160} governance and not known admin addresses. 

So, totally out of 7,470 proxy contract, \textbf{3558} are controlled by an EOA address, \textbf{988} are controlled by a multi signature wallet and \textbf{2924} addresses are governance controlled or our methodology could not find their type.

\clearpage
\section{Attacking Universal Upgradeable Proxy Standard (UUPS) contracts} 
\label{sec:attackUUPS}

In Section~\ref{sysComplexity}, we discussed that one of the main challenges to the \texttt{DELEGATECALL}-based data separation pattern (Section~\ref{sec:delegatecall}) is that the constructor inside the implementation contract cannot initialize the proxy itself. So instead, there should be a regular function inside the implementation contract named \textit{Initialize} function that can be called just once after deployment by the proxy contract and has the same functionality as the constructor function. Therefore, the contract creator must call the initialize function quickly after deploying the proxy contract. The Initialize function does not have any access control because it is considered to be called once, and this function is responsible for defining the owner. So, before calling this function, the owner's address is not set, and there is no way to have an access control check for the sender. This is why there should be a check to ensure that this function can just be called once at deployment and not after. 
The proxy contract deployer will define and initialize the address of the contract owner via the initialize function. So if the deployer forgets to initialize the contract, any external address can call initialize function and change the owner of the contract to her desired address, and take control of the contract.

The \texttt{Initialize} function can also be called from the implementation contract itself (instead of calling the function by proxy contract). This call will alter the storage of the implementation contract and not the proxy contract. Suppose the deployer forgets to call the \texttt{Initialize} function directly from the implementation contract. Any malicious address can call this function from the implementation contract and change the owner inside the implementation contract, and take control of this contract. 
This malicious actor can change the storage of the implementation contract by calling functions inside it. However, it is not a risk to the system because the proxy contract is responsible for keeping the data in this system and not implementing it. There should be a risk to the system if this malicious actor can change the logic of the implementation contract or self-destruct it. Changing the logic of the implementation contract is not doable in typical cases because implementation contracts are not supposed to be upgradeable. However, there are ways to self-destruct the implementation contract. 

There are two main ways to self-destruct a contract: 1)if the implementation contract has \texttt{SELFDESTRUCT} inside its logic and by calling it. 2)Having a \texttt{DELEGATECALL} or \texttt{CALLCODE} to another contract that has \texttt{SELFDESTRUCT} logic inside~\cite{frowisnot}.

So, we should check if the implementation contract uses \texttt{SELFDESTRUCT} or has \texttt{DELEGATECALL} or \texttt{CALLCODE} to an address that a malicious party can control. If there is a way, the malicious party can self-destruct the implementation contract, and all calls to the proxy will fail. It is a Denial of Service (DoS) attack on the Dapp.
If the Dapp has an upgrade function inside its proxy contract, then the admin of the proxy contract can upgrade into a new version of the implementation contract. This attack was explained in December 2020 by  Trail of Bits team when they audited the code of Aave, a lending project~\cite{aaveBreak}.

Nevertheless, what if the upgrade function is inside the implementation contract and not the proxy contract. As mentioned in Section~\ref{sec:delegatecall}, in UUPS upgradeable contracts, the upgrade function resides in the implementation contract. So there is no way to upgrade the system by the proxy contract. Therefore, if an attacker takes control of the implementation contract by calling the initialize function directly from the implementation contract and then self-destruct it, there is no way to upgrade it. Consequently, the proxy will be locked forever. 
All UUPS contracts that used the Openzepplin UUPS library, whose implementation contact is not initialized, are susceptible to this attack. Because there is a function in the implementation contract of this library named \textit{upgradeToAndCall}, in which the owner can change a target address and then delegate call into the newly changed target address. This attack vector was found in September 2021 and announced by OpenZepplin team~\cite{securityAdvise,uupsAttacks}. There is an easy way to mitigate this attack by calling the initialize function directly from the implementation contract. 

We try to check all UUPS contracts that we find in Section~\ref{sec:proxyFinding} that if any of them can be exploited in this way. We check all of them manually, and the method of checking them is described below:

\begin{enumerate}
 \item Check if the implementation contract is not initialized
 \item Find initialize function inside the implementation contract
 \item check if anybody can call this initialize function directly from the implementation contract and change the owner of the contract
 \begin{itemize}
 \item Filter those that have a modifier that blocks direct calls to the implementation contract (there is a modifier that just let transactions that come from the proxy contract and blocks direct calls to the implementation contract itself)
 \end{itemize}
 \item Check if there is a way inside the implementation to self-destruct 
 \item Check if there is a function in the implementation contract which has a delegate call to a target address
 \item Check if the target address is changeable by a malicious actor
\end{enumerate}

After reviewing the list above, we found 15 contracts in our data set that were exploitable until September 9, 2021. The openzeppelin team patched them by initializing the contract. An attacker could deploy a new malicious contract that executes self-destruct on any calls to it. Then take control of the implementation contract by calling the initialize function them. Afterward, the attacker finds the function inside the implementation contract with a delegate call inside it and finds the target address. There should be a function inside the implementation contract to change the address to the malicious contract that the attacker deployed recently. The attacker calls the function to execute a delegate call into the malicious contract and then self-destruct the implementation contract.

We find 61 UUPS contracts that are not initialized, and anybody can take control of these implementation contracts. However, because these contracts do not use delegate calls or self-destruct, they are not exploitable by this type of attack.

%% file: main.bbl
\begin{thebibliography}{10}
\providecommand{\url}[1]{\texttt{#1}}
\providecommand{\urlprefix}{URL }
\providecommand{\doi}[1]{https://doi.org/#1}

\bibitem{bentFinanceHack}
Bent update. Tech. rep., Bent Finance,
  \url{https://bentfi.medium.com/bent-update-12ae69a41dc6}

\bibitem{aaveBreak}
Breaking aave upgradeability. Tech. rep., Trail of Bits,
  \url{https://blog.trailofbits.com/2020/12/16/breaking-aave-upgradeability/}

\bibitem{tobBlogPost}
Contract upgrade anti-patterns. Tech. rep., Trail of Bits,
  \url{https://blog.trailofbits.com/2018/09/05/contract-upgrade-anti-patterns/}

\bibitem{certikReport}
The state of defi security 2021. Tech. rep., Certik Company,
  \url{https://blog.openzeppelin.com/the-state-of-smart-contract-upgrades/}

\bibitem{delegatecallForwarders}
Buterin, V.: Delegatecall forwarders: how to save ~50-98
  contracts with the same code
  \url{https://www.reddit.com/r/ethereum/comments/6c1jui/delegatecall_forwarders_how_to_save_5098_on/}

\bibitem{chen2021smart}
Chen, J., Xia, X., Lo, D., Grundy, J.: Why do smart contracts self-destruct?
  investigating the selfdestruct function on ethereum. ACM Transactions on
  Software Engineering and Methodology (TOSEM)  \textbf{31}(2),  1--37 (2021)

\bibitem{chen2017adaptive}
Chen, T., Li, X., Wang, Y., Chen, J., Li, Z., Luo, X., Au, M.H., Zhang, X.: An
  adaptive gas cost mechanism for ethereum to defend against under-priced dos
  attacks. In: International {C}onference on {I}nformation {S}ecurity practice
  and experience. pp. 3--24. Springer (2017)

\bibitem{dhillon2017dao}
Dhillon, V., Metcalf, D., Hooper, M.: The {DAO} hacked. In: Blockchain Enabled
  Applications, pp. 67--78. Springer (2017)

\bibitem{frowisnot}
Fr{\"o}wis, M., B{\"o}hme, R.: Not all code are {Create2} equal
  \url{https://informationsecurity.uibk.ac.at/pdfs/FB-Ethereum-Create2.pdf}

\bibitem{he2020characterizing}
He, N., Wu, L., Wang, H., Guo, Y., Jiang, X.: Characterizing code clones in the
  ethereum smart contract ecosystem. In: International Conference on Financial
  Cryptography and Data Security. pp. 654--675. Springer (2020)

\bibitem{mccorry2021sok}
McCorry, P., Buckland, C., Yee, B., Song, D.: Sok: Validating bridges as a
  scaling solution for blockchains. Cryptology ePrint Archive  (2021)

\bibitem{minimalProxy}
Murray, P., Welch, N., Messerman, J.: Minimal proxy contract. EIP-1167 (2018)

\bibitem{smart_contract_sanctuary}
Ortner, M., Eskandari, S.: Smart contract sanctuary
  \url{https://github.com/tintinweb/smart-contract-sanctuary}

\bibitem{securityAdvise}
Palladino, S.: Security advisory: Initialize {UUPS} implementation contracts
  \url{https://forum.openzeppelin.com/t/security-advisory-initialize-uups-implementation-contracts/15301}

\bibitem{openzeppelinPost}
PALLADINO, S.: The state of smart contract upgrades
  \url{https://blog.openzeppelin.com/the-state-of-smart-contract-upgrades/}

\bibitem{uupsAttacks}
Palladino, S.: Uupsupgradeable vulnerability post-mortem
  \url{https://forum.openzeppelin.com/t/uupsupgradeable-vulnerability-post-mortem/15680}

\bibitem{perez2022dissimilar}
Perez, D., Gudgeon, L.: Dissimilar redundancy in defi. arXiv preprint
  arXiv:2201.12563  (2022)

\bibitem{perez2019broken}
Perez, D., Livshits, B.: Broken metre: Attacking resource metering in evm.
  arXiv preprint arXiv:1909.07220  (2019)

\bibitem{pinna2019massive}
Pinna, A., Ibba, S., Baralla, G., Tonelli, R., Marchesi, M.: A massive analysis
  of {E}thereum smart contracts empirical study and code metrics. IEEE Access
  \textbf{7},  78194--78213 (2019)

\bibitem{reijsbergen2021transaction}
Reijsbergen, D., Sridhar, S., Monnot, B., Leonardos, S., Skoulakis, S.,
  Piliouras, G.: Transaction fees on a honeymoon: Ethereum's eip-1559 one month
  later. In: 2021 IEEE International Conference on Blockchain (Blockchain). pp.
  196--204. IEEE (2021)

\bibitem{rodler2021evmpatch}
Rodler, M., Li, W., Karame, G.O., Davi, L.: $\{$EVMPatch$\}$: Timely and
  automated patching of ethereum smart contracts. In: 30th USENIX Security
  Symposium (USENIX Security 21). pp. 1289--1306 (2021)

\bibitem{victor2019measuring}
Victor, F., L{\"u}ders, B.K.: Measuring ethereum-based erc20 token networks.
  In: International Conference on Financial Cryptography and Data Security. pp.
  113--129. Springer (2019)

\bibitem{walch2016path}
Walch, A.: The path of the blockchain lexicon (and the law). Rev. Banking \&
  Fin. L.  \textbf{36}, ~713 (2016)

\end{thebibliography}
